\documentclass{mem}
\usepackage{natbib}\usepackage{txfonts}\usepackage{balance}
\usepackage{graphicx}
\usepackage[a4paper]{hyperref}
\idline{75}{282}
\begin{document}
\def\teff{$T\rm_{eff }$}
\def\kms{$\mathrm {km s}^{-1}$}

\title{
Completing the Census of (Bright) Variable Stars in Galactic Globular Clusters
}

   \subtitle{}

\author{
M. \,Catelan,\inst{1} 
H. A. Smith,\inst{2} 
B. J. Pritzl,\inst{3}
J. Borissova,\inst{4}
C. Cacciari,\inst{5}
R. Contreras,\inst{1}
T. M. Corwin,\inst{6}
N. De Lee,\inst{2}
M. E. Escobar,\inst{1}
A. C. Layden,\inst{7}
C. Navarro,\inst{1}
G. Prieto,\inst{1}
R. Salinas,\inst{1,8}
P. B. Stetson,\inst{9}
A. V. Sweigart,\inst{10}
E. Vidal,\inst{1}
M. Zoccali,\inst{1}
\and
M. Zorotovic\inst{1}
          }

  \offprints{M. Catelan}

\institute{
Pontificia Universidad Cat\'olica de Chile, Departamento de 
Astronom\'\i a y Astrof\'\i sica, Av. Vicu\~{n}a Mackenna 4860, 
782-0436 Macul, Santiago, Chile;
\email{mcatelan@astro.puc.cl}
\and
Dept.\ of Physics and Astronomy, Michigan State Univ., East Lansing, 
MI 48824, USA
\and
Macalester College, 1600 Grand Avenue, Saint Paul, MN 55105, USA 
\and
European Southern Observatory, Av. Alonso de C\'ordova 3107, 763-0581 
Vitacura, Santiago, Chile
\and
INAF -- Osservatorio Astronomico di Bologna, via Ranzani 1, I-40127 
Bologna, Italy 
\and
Dept. of Physics, University of North Carolina at Charlotte, 
Charlotte, NC 28223, USA
\and
Dept.\ of Physics and Astronomy, 104 Overman Hall, Bowling Green State
University, Bowling Green, OH 43403, USA
\and
Grupo de Astronom\'\i a, Facultad de Ciencias F\'\i sicas y Matem\'aticas, 
Universidad de Concepci\'on, Concepci\'on, Chile
\and
Dominion Astrophysical Observatory, Herzberg Institute of Astrophysics,
National Research Council, 5071 West Saanich Road, Victoria, BC V9E 2E7, 
Canada
\and
NASA Goddard Space Flight Center, Exploration of the Universe
Division, Code 667, Greenbelt, MD 20771, USA
}

\authorrunning{Catelan et al.}

\titlerunning{Variable Stars in Globular Clusters}

\abstract{
We present a long-term project aimed at completing the census 
of (bright) variable stars in Galactic globular clusters. While our 
main aim is to obtain a reliable assessment of the populations of 
RR Lyrae and type II Cepheid stars in the Galactic globular cluster 
system, due attention is also being paid to other types of variables,  
including SX Phoenicis stars, long-period variables, and eclipsing binaries. 
}
\maketitle{}

\section{Introduction}

Variable stars remain one of the most important types of object 
in astronomy. They are widely used in a variety of different areas, 
from tests of stellar structure and evolution theory to the 
determination of the extragalactic distance scale. Globular 
star clusters present a particularly interesting ``astrophysical 
laboratory'' for the study of variable stars, since in these 
environments one is offered the chance of studying statistically 
significant samples of variable stars in different evolutionary 
stages, all at the same distance from us, with a common age, and 
(nearly) the same chemical composition.  

By the early-1990's, it was widely perceived that ``most 
variables that are in [globular] clusters have by now become 
discovered'' \citep{nsea91}. In the \citeauthor{nsea91} compilation, 
only a few ``notable exceptions'' were mentioned.
More 
specifically, \citet{nsea91} estimate that ``only 6\% of the 
cluster [RR Lyrae] variables remain to be discovered.'' Has this 
assessment withstood the test of time? 

Unfortunately, not quite: some 15 years later, it is now clear 
that not only do several of the ``notable exceptions'' mentioned 
by \citet{nsea91} remain 
poorly studied in terms of variability, but also, and 
importantly, many---if not all---globular 
clusters whose variable star populations were considered 
exhaustively known by the early 1990's are now known to 
be affected by severe incompleteness in their reported numbers 
of even the brighter variable stars (such as the RR Lyrae). 

Consider, as an example, the case of M3 (NGC~5272), first studied 
by \citet{sb13}, and subsequently investigated in detail 
by \citet{ecea98}, \citet{jkea98}, \citet{cc01}, and \citet{jsea02},
among others. According to the compilation of 
variable stars in globular clusters by \citet{ccea01}, by the end
of the 1990's M3 had a total of 182 catalogued RR Lyrae variables. 
By the year 2004, about 230 RR Lyrae variables were already known 
in the cluster \citep{gbea00,gcea04}---thus representing an increment 
of 26\% with respect to the \citeauthor{ccea01} compilation. 

There are two main reasons for the somewhat unexpected increase in 
the numbers of (bright) variable stars in globular clusters. First, 
the pre-1990's studies were primarily based on photographic 
photometry, which in many cases appears not to have been precise 
enough to tell small-amplitude variables from noise. Second, it
was only very recently that an image-subtraction technique was 
developed which is capable of quickly, efficiently and automatically 
subtracting out the constant information from one CCD frame of a 
cluster to the next, also matching the differences in seeing between 
the frames in the process. This remarkable achievement is due 
primarily to tools created by \citet{al98} and 
\citet{ca00}, which have become accessible to the astronomical 
community in the form of the ISIS image-subtraction program. 

While very efficient in {\em detecting} variable stars in crowded 
fields, ISIS presents the drawback of providing light curves in 
flux values relative to a reference image only. For this reason, 
ISIS does not provide light curves in standard magnitudes, and the 
reference image has to be processed independently for this purpose. 
While DAOPHOT/ALLFRAME \citep{pbs94} offer excellent tools to 
perform absolute photometry in the crowded regions found in 
globular clusters, it is still often the case that the 
variable stars cannot have their absolute fluxes reliably measured 
in the ISIS reference image, making it very difficult to convert the 
ISIS relative-flux light curves into calibrated ones without 
additional images obtained with higher spatial resolution. 

In the present paper, we present our long-term project to complete 
the census of (bright) variable stars in Galactic globular clusters. 
The tools adopted to analyze our time-series photometry are ISIS 
(detection, light curves in relative fluxes for period and 
variability type determination) and DAOPHOT/ALLFRAME (light curves 
in standard magnitudes, positioning of the variables in the 
color-magnitude diagram, determination of variability types, and 
determination of physical properties on the basis of the Fourier 
decomposition parameters).

\section{Observational Data}
The data upon which this project is based comes primarily from 
small (i.e., 2m class or less) telescopes, such as the Danish 1.54m, 
the CTIO 0.9m and 1.3m, the LCO Warsaw 1.3m, and the Rozhen 2m 
telescopes. Data available in public archives are also being used. 
The observational data, generally consisting of 
sets of $B$, $V$, and (less often) $I$ images, are usually collected 
during observing runs split in blocks of several days, spread over 
several months, which helps ensure proper coverage of the periods 
of both RR Lyrae stars and type II Cepheids. In the case of the 
CTIO data, most of our data has been obtained in 
service mode, with one or more datapoints being collected per 
night over a period of several months.  

In what follows, we describe some of the latest results of this 
project, which have not yet appeared in the refereed literature.

   \begin{figure}[t]
   \centering
   \includegraphics[width=5.5cm]{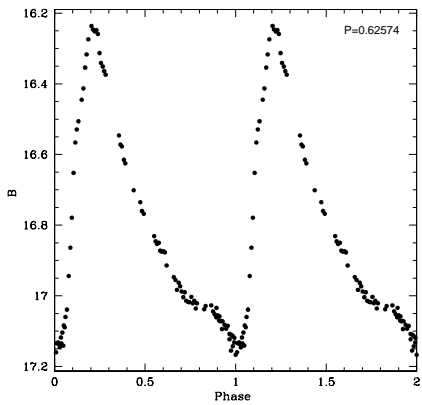}
      \caption{Light curve in $B$ for a newly discovered 
               RRab Lyrae variable in M69.  
              }
         \label{Fig1}
   \end{figure}

   \begin{figure}
   \centering
   \includegraphics[width=5.5cm]{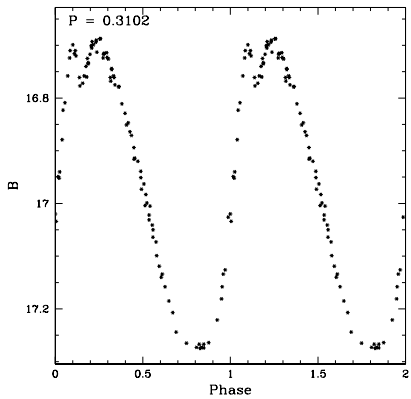}
      \caption{Light curve in $B$ for a newly discovered 
               RRc Lyrae variable in NGC~5286.  
              }
         \label{Fig2}
   \end{figure}

   \begin{figure}[t]
   \centering
   \includegraphics[width=5.5cm]{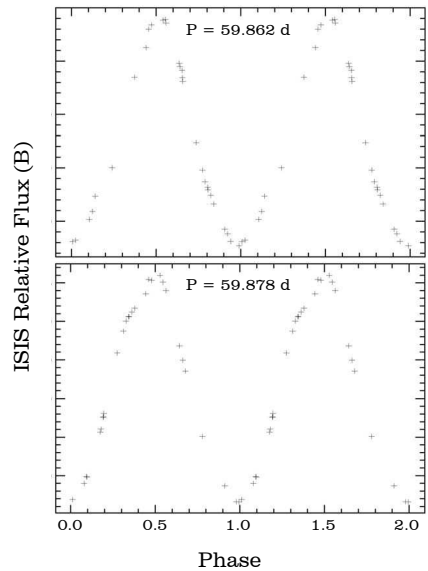}
      \caption{Light curves for two newly discovered type II 
               Cepheids in M28. 
              }
         \label{Fig3}
   \end{figure}

   \begin{figure}[h]
   \centering
   \includegraphics[width=5.5cm]{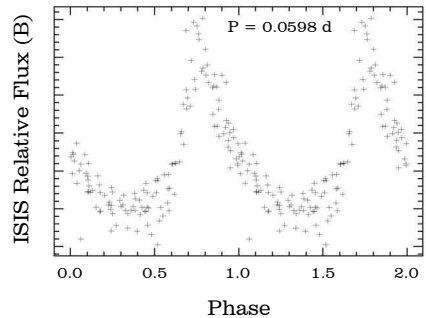}
      \caption{Light curves for a newly discovered SX Phe 
               candidate in NGC~2808. 
              }
         \label{Fig4}
   \end{figure}

\section{Some Recent Results}

\subsection{The Metal-Rich Globular Cluster M69 (NGC~6637)}

This project is based on data collected at LCO with the Warsaw
1.3m telescope. In Figure~1 is shown a light curve of an RR Lyrae 
star that was discovered in the field of the cluster in the course 
of our study. Its period is relatively long for a metal-rich 
globular cluster, thus bringing to mind the cases of V9 in 
47~Tucanae and the many long-period RR Lyrae variables 
in NGC~6388 and NGC~6441. While
membership status is unclear at present, the star does appear to 
lie sufficiently close to the cluster's horizontal branch (HB) in 
the color-magnitude diagram as to merit further analysis of this
possibility. De Lee et al. (these proceedings) discuss the 
possibility that NGC~6304 may provide yet another example of a 
metal-rich globular cluster with long-period RR Lyrae variables.

\subsection{NGC~5286: an RR Lyrae-Rich Globular in the Canis Major 
   Dwarf Spheroidal Galaxy (CMa dSph)?}

This project is also based on data collected at the LCO with the 
Warsaw 1.3m telescope. NGC~5286 is a particularly interesting 
cluster due to its suggested association with the CMa dSph 
\citep{pfea04,nfmea04}. 
We have found a rich 
harvest of RR Lyrae stars in this cluster, increasing quite 
substantially the number of known variables. 
Figure~2 shows a light curve for a newly discovered RRc star.

\subsection{Type II Cepheids in the Blue HB Globular Cluster M28 
            (NGC~6626)}

The data used in this project were collected using the CTIO 1.3m 
telescope in service mode. 
A fairly large number of variable stars was discovered, 
including short-period ($P < 0.3$~d) RRc variables; long-period
($P > 0.7$~d) RRab variables; type II Cepheids (Fig.~3), 
including four stars with periods very close to 60~d; 
and long-period or 
semi-regular variables with periods longer than 50~d.

\subsection{SX Phoenicis Variables in NGC~2808}

Until recently, NGC~2808 had been though to be essentially 
devoid of RR Lyrae variables. In the course of the present project, 
however, we were able to discover a sizeable number of RR Lyrae 
stars in the cluster, as described in \citet{mcea04}. Recently, 
we have also found that the cluster contains a significant 
number of SX Phe variables, and a sample light curve is provided 
in Figure~4.

\section{Conclusions}
This project is still in its beginning, and it is already clear that 
there is a long road before we can consider the variable star 
populations in globular clusters as well known. In addition, while 
our attention (and observing strategies) have mainly focused on 
the brighter RR Lyrae and type II Cepheid variables, it is becoming
increasingly clear that there is also a rich harvest of SX Phe 
and eclipsing variable stars in globular clusters waiting to be 
found. It is our hope that the astronomical community will continue 
to support the operation of 1m-class telescopes throughout the world, 
so that the study of variable stars---upon which so much of modern 
astrophysics relies---can truly become a well-established enterprise.

\begin{acknowledgements}
Support for M.C., M.E.E., R.C., C.N., G.P., and R.S. 
was provided by Fondecyt \#1030954; for H.A.S., 
by NSF grant AST-0205813; and for B.J.P., by 
CAREER award AST 99-84073.
\end{acknowledgements}

\bibliographystyle{aa}

\begin{thebibliography}{}

\bibitem[Alard(2000)]{ca00}
  Alard, C. 2000, \aaps, 144, 363

\bibitem[Alard \& Lupton(1998)]{al98}
  Alard, C., \& Lupton, R. H. 1998, \apj, 503, 325

\bibitem[Bailey(1913)]{sb13}
  Bailey, S. I. 1913, Harv. Coll. Observ. Annals, 78, 1

\bibitem[Bakos et al.(2000)]{gbea00}
  Bakos, G. \'{A}, Benk\"{o}, J. M., \& Jurcsik, J. 2000, AcA, 50, 221

\bibitem[Carretta et al.(1998)]{ecea98}
  Carretta, E., Cacciari, C., Ferraro, F. R., Fusi Pecci, F., \& 
    Tessicini, G. 1998, \mnras, 298, 1005


\bibitem[Clement et al.(2001)]{ccea01}
  Clement, C. M., et al. 2001, \aj, 122, 2587

\bibitem[Clementini et al.(2004)]{gcea04}
  Clementini, G., Corwin, T. M., Carney, B. W., \& Sumerel, A. N. 2004, \aj, 
  127, 938

\bibitem[Corwin \& Carney(2001)]{cc01}
  Corwin, T. M., \& Carney, B. W. 2001, \aj, 122, 3183

\bibitem[Corwin et al.(2004)]{mcea04}
  Corwin, T. M., Catelan, M., Borissova, J., \& Smith, H. A. 2004, \aap, 421, 667

\bibitem[Frinchaboy et al.(2004)]{pfea04}
  Frinchaboy, P. M., Majewski, S. R., Crane, J. D., Reid, I. N., Rocha-Pinto, 
  H. J., Phelps, R. L., Patterson, R. J., \& Mu\~noz, R. R. 2004, \apjl, 602, L21

\bibitem[Kaluzny et al.(1998)]{jkea98}
  Kaluzny, J., Hilditch, R. W., Clement, C., \& Rucinski, S. M. 1998, 
    \mnras, 296, 347

\bibitem[Martin et al.(2004)]{nfmea04}
  Martin, N. F., Ibata, R. A., Bellazzini, M., Irwin, M. J., Lewis, G. F., \& 
  Dehnen, W. 2004, \mnras, 348, 12

\bibitem[Stetson(1994)]{pbs94}
  Stetson, P. B. 1994, \pasp, 106, 250

\bibitem[Strader et al.(2002)]{jsea02}
  Strader, J., Everitt, H. O., \& Danford, S. 2002, \mnras, 335, 621

\bibitem[Suntzeff et al.(1991)]{nsea91}
  Suntzeff, N. B., Kinman, T. D., \& Kraft, R. P. 1991, \apj, 367, 528


\end{thebibliography}

\end{document}